\documentclass[]{aa}  
\usepackage{amsmath}
\usepackage{bm}
\usepackage{makecell}
\usepackage{booktabs}
\usepackage{multirow}
\usepackage{graphicx}
\usepackage[varg]{txfonts}
\usepackage{hyperref}
\usepackage{xcolor}
\usepackage{ulem}
\hypersetup{
    colorlinks=true,
    citecolor=blue,
    linkcolor=red,
    filecolor=magenta,
    urlcolor=cyan,
} 
\bibpunct{(}{)}{;}{a}{}{,}
\begin{document}   
\title{Confirmation of TiO absorption and tentative detection of MgH and CrH in the atmosphere of HAT-P-41b}

\author{
    C. Jiang\inst{1, 2}
    \and G. Chen\inst{1, 3}
    \and F. Murgas\inst{4,5}
    \and E. Pall\'e\inst{4,5}
    \and H. Parviainen\inst{4,5}
    \and Y. Ma\inst{1}}

\institute{
    CAS Key Laboratory of Planetary Sciences, Purple Mountain Observatory, Chinese Academy of Sciences, No. 10 Yuanhua Road, Qixia District, 210023 Nanjing, PR China \\
    \email{guochen@pmo.ac.cn}
    \and School of Astronomy and Space Science, University of Science and Technology of China, No. 96 Jinzhai Road, Baohe District, 230026 Hefei, PR China
    \and CAS Center for Excellence in Comparative Planetology, No. 96 Jinzhai Road, Baohe District, 230026 Hefei, PR China
    \and Instituto de Astrof\'isica de Canarias (IAC), V\'ia L\'actea s/n, 38205 La Laguna, Tenerife, Spain
    \and Departamento de Astrof\'isica, Universidad de La Laguna (ULL), C/ Padre Herrera, 38206 La Laguna, Tenerife, Spain}

\date{Received ... Accpected ...}

\abstract
{
Understanding the role of optical absorbers is critical for linking the properties of the day-side and terminator atmospheres of hot Jupiters. This study aims to identify the signatures of optical absorbers in the atmosphere of the hot Jupiter HAT-P-41b. We conducted five transit observations of this planet to obtain its optical transmission spectra using the Gran Telescopio Canarias (GTC). We performed atmospheric retrievals assuming free abundances of 12 chemical species. Our Bayesian model comparisons revealed strong evidence for TiO absorption ($\Delta\ln \mathcal{Z}=21.02$), modest evidence for CrH ($\Delta\ln \mathcal{Z}=3.73$), and weak evidence for MgH ($\Delta\ln \mathcal{Z}=2.32$). When we combined the GTC transmission spectrum with previously published Hubble Space Telescope (HST) and Spitzer data, the retrieval results and model inferences remained consistent. In conclusion, HAT-P-41b has a metal-rich atmosphere with no high-altitude clouds or hazes. Further observations of its day-side atmosphere should be made to confirm the hints of a thermal inversion in the upper atmosphere suggested by our results.
}


\maketitle


\section{Introduction}
Accurate characterization of exoplanet atmospheres is crucial in solving the puzzle of planetary formation and evolution. Precise spectroscopic measurements of exoplanet atmospheres, available with current observational instruments, are mainly focused on close-in giant planets, including hot Jupiters. Certain metals, metal oxides and hydrides have very strong opacity in the optical wavelengths \citep{2007ApJS..168..140S}, which can absorb the stellar radiation and heat the upper atmosphere, leading to a thermal inversion in the temperature-pressure (T-P) profile and thermal emission features in the emission spectrum \citep{2008ApJ...678.1436B, 2010A&A...520A..27G}. While it is difficult to detect optical absorbers in the emission spectrum, alkali metals (e.g., Na and K), metal oxides (e.g., TiO and VO), and hydrides (e.g., CrH and FeH) are expected to exhibit strong absorption features in the optical band of transmission spectrum \citep{2000ApJ...537..916S, 2008ApJ...678.1419F, 2010ApJ...709.1396F}. 
 
The hot Jupiter HAT-P-41b, discovered by \cite{2012AJ....144..139H}, is an intriguing target for further study due to its highly inflated nature. The planet orbits an F-type star ($T_{\rm eff}=6390~{\rm K}$, $M_\star\sim1.4M_\odot$, $R_\star\sim1.7~R_\odot$, $\rm [Fe/H]=0.21$) every 2.7 days and has a mass of $\sim$$0.8~M_{\rm J}$, a radius of $\sim$$1.7~R_{\rm J}$, and an equilibrium temperature of $1941~{\rm K}$. \cite{2020AJ....159..204W} presented a 0.2 -- 0.8 $\rm\mu m$ transmission spectrum using the G280 grism of HST WFC3 UVIS, which revealed a significant scattering slope and strong absorption features of TiO and/or VO. \citet{2020ApJ...902L..19L} performed retrievals on the 0.2--5.0~$\mu$m combined spectrum, consisting of HST WFC3 (G280, G141) and Spitzer, and found a potential combination of H$^-$, CrH, AlO, VO, and Na to explain the ultraviolet to optical wavelengths. However, the presence of $\rm H^-$ in the planet's terminator atmosphere remains inconclusive (\citealt{2023AJ....165..112W}). \cite{2021AJ....161...51S} conducted a comprehensive retrieval analysis of the 0.3--5.0 $\mu$m spectrum derived from HST STIS, HST WFC3, and Spitzer transit observations and found that the atmosphere of HAT-P-41b is cloud-free, haze-free, and metal-rich, with distinct water vapor absorption and tentative optical absorption. The day-side upper atmosphere of HAT-P-41b does not exhibit strong evidence for a thermal inversion, as suggested by the emission spectrum analyses \citep{2021NatAs...5.1224M, 2022AJ....163..190F, 2022ApJS..260....3C}. The observed transmission spectrum suggest that optical absorbers such as metal oxides and hydrides may be present in the planet's terminator atmosphere, but the emission spectrum did not reveal a distinct thermal inversion in the day-side atmosphere. Further confirmation and identification of the candidate optical absorbers in the atmosphere of HAT-P-41b is needed to resolve this discrepancy.

Our goal is to further investigate the atmosphere of HAT-P-41b using low-resolution transmission spectrum. Multiple transit observations with the GTC OSIRIS instrument \citep{2000SPIE.4008..623C} should enable us to obtain the optical transmission spectrum with higher photometric precision and spectral resolution than the HST WFC3 UVIS and STIS instruments. By performing atmospheric retrievals and Bayesian model comparisons, we aim to obtain more precise estimates of the abundances of the chemical species, better constrain previously reported optical absorbers, and possibly discover new signatures in the atmosphere of HAT-P-41b.

This paper is organized as follows. We summarize the observations and the data reduction procedures in the next section. Section \ref{sect:lightcurve} illustrates the analysis of the GTC transit light curves. The atmospheric retrievals are demonstrated in Sect. \ref{sect:retrievals}. We further discuss the joint retrievals and then draw our conclusion in Sect. \ref{sect:dis_and_con}. 

\section{Observations and data reduction} \label{Observations}

We observed five transits of HAT-P-41b using the GTC OSIRIS instrument between 2013 and 2022 (hereafter OB-1 to OB-5). The observation summary is listed in Table \ref{table:observation}. The instrument OSIRIS has an unvignetted field of view of $7.4'$ and a pixel scale of $0.254''$. Its detector consists of a mosaic of two Marconi CCDs ($1024\times2048$ pixels) with a 37-pixel ($9.4''$) gap between them after a $2\times2$ pixel binning. The observations were performed in the longslit spectroscopic mode with the R1000R grism covering a wavelength range of 5100 -- 10\,000~\AA. The gap between the CCDs was parallel to the dispersion direction and thus would not affect the spectra. We adopted a $12''$-wide slit in OB-1, OB-3, and OB-4, and a $40''$-wide slit in OB-2 and OB-5 to avoid flux loss due to the partial coverage of the point spread function (PSF). A comparison star (2MASS 19493265$+$0439270 for OB-1, OB-3, OB-4, and OB-5; 2MASS 19494153$+$0437350 for OB-2; \citealt{2003yCat.2246....0C}) was simultaneously observed with the target star through the slit in the five observations.

The data were reduced following the methods described in our previous work \citep{2022A&A...664A..50J} to produce the spectroscopic light curves with 5-nm passbands from 521.8~nm to 922.4~nm, where the passband of 756.8 -- 767.4~nm covering the telluric oxygen A-band was discarded in subsequent analysis due to the very-low signal-to-noise ratio. The broadband light curves (i.e., white-light curves) were obtained in the same manner but integrated from 521.8~nm to 922.4~nm excluding the oxygen A-band.

\section{Light curve analysis} \label{sect:lightcurve}

\subsection{Transit model}

We employed the Python package \texttt{PyTransit} \citep{2015MNRAS.450.3233P} to model the transit light curves, assuming circular orbits. There are five transit parameters in the model: the planet-to-star radius ratio ($R_{\rm p}/R_{_s}$), normalized orbital semimajor axis ($a/R_\star$), orbital inclination ($i$), central transit epoch ($t_{\rm c}$), and orbital period ($P$). For each transit, the light curves were fit separately. 

For broadband light curve fitting, we employed a four-parameter nonlinear stellar limb darkening law. The limb darkening coefficients ($u_1$ to $u_4$) were derived using \texttt{LDTK} \citep{2015MNRAS.453.3821P} and kept fixed in light curve fitting. The orbital period was determined based on the transit ephemeris updated by \cite{2020AJ....159..204W}, with the central epoch of each transit being a free parameter under a uniform prior. 
To avoid systematic biases in the transmission spectra and to facilitate subsequent joint atmospheric retrievals with the transmission spectra provided by \cite{2021AJ....161...51S}, we adopted the orbital semimajor axis and inclination values consistent with those used in \cite{2021AJ....161...51S}.
Furthermore, we used the broadband radius ratio measured with HST STIS G750L ($R_{\rm p}/R_{\rm s}=0.10159\pm0.00042$, \citealt{2021AJ....161...51S}), which essentially matches the wavelength coverage of OSIRIS R1000R, as the Gaussian prior for the broadband radius ratio. This was done to ensure consistency across the five transit observations and to obtain precise central transit epochs and broadband systematic noise.
Table \ref{table:input} shows the adopted parameters of the HAT-P-41 system.

We performed separate fittings for the narrowband light curve in each passband. The parameter settings for the narrowband fitting were kept consistent with those of the broadband, with the exception of $R_{\rm p}$ and $T_{\rm c}$. The radius ratio was treated as a free parameter under a uniform prior, while the central epoch of each transit was fixed to the best-fit value derived from the broadband light curve fitting. 
To mitigate the impact of wavelength-independent systematic noise (i.e., common-mode systematics) and enhance the precision of the derived transmission spectra, we employed the common-mode correction method outlined in \cite{2013MNRAS.428.3680G}. This involved dividing the best-fit broadband light curve by the best-fit transit model to obtain the common-mode systematics. Each narrowband light curve was then divided by the common-mode systematics to apply this correction. We note that \cite{2013MNRAS.428.3680G} used the GP systematics plus one to approximate the common mode systematics, while we used the best-fit broad-band light curve divided by the best-fit transit model. Given that the amplitudes of the GP systematics were much smaller than one, these two approximations are virtually indistinguishable.

\begin{table*}
\caption{Input parameters of the HAT-P-41 system used in this work.} 
\label{table:input} 
\centering 
\begin{tabular}{l l c c} 
\hline\hline 
Symbols & Parameters & Values & References\\ 
\hline 
Stellar parameters \\
$T_{\rm eff}$ & Stellar effective temperature (K) & $6390 \pm 100$ & [1] \\
$\rm [Fe/H]$ & Stellar metallicity & $0.21 \pm 0.10$ & [1] \\
$\log_{10} g_\star$ & Stellar gravity (cgs) & $4.14 \pm 0.02$ & [1] \\
$M_\star$ & Stellar mass ($M_\sun$) & $1.418 \pm 0.047$ & [1] \\
$R_\star$ & Stellar radius ($R_\sun$) & $1.683 ^{+0.058}_{-0.036} $ & [1] \\

\\ Planetary parameters \\
$M_{\rm p}$ & Planetary mass ($M_{\rm J}$) & $0.800 \pm 0.102$ & [1] \\
$R_{\rm p}$ & Planetary radius ($R_{\rm J}$) & $1.685 \pm 0.076$ & [1] \\
$T_{\rm eq}$ & Planetary equilibrium temperature (K) & $1941 \pm 38$ & [1] \\
$P$ & Orbital period (day) & 2.69404861 & [2] \\
$t_0$ & Initial transit epoch ($\rm BJD_{TDB}$) & 2456600.29325 & [2] \\
$a/R_\star$ & Orbital semi-major axis relative to the stellar radius & $5.44 \pm 0.15$ & [1] \\
$i$ & Orbital inclination (deg) & $87.7 \pm 1.0$ & [1] \\

$R_{\rm p}/R_{\rm s}$ & Planet-to-star radius ratio (HST STIS G750L) & $0.10159 \pm 0.00042$ & [3] \\

\hline 
\end{tabular}

\tablebib{
    [1] \cite{2012AJ....144..139H}
    [2] \cite{2020AJ....159..204W}
    [3] \cite{2021AJ....161...51S}
}
    
\end{table*}

\subsection{Correction for the companion star contamination}
\label{ap:companion_correction}

\cite{2012AJ....144..139H} reported a resolved neighboring star, located $\sim$3.56\arcsec away and 3.5~mag fainter than HAT-P-41 in the $i$ band. \cite{2015A&A...575A..23W} further observed, through lucky imaging, that this neighboring star has magnitude contrasts of $\Delta i'=3.65 \pm 0.05$ and $\Delta z'=3.40 \pm 0.05$, and a spectral type between F5 and M2 assuming physical companionship. According to the Gaia observations (Table \ref{table:companion}; \citealt{2016A&A...595A...1G, 2023A&A...674A...1G}), these two stars share similar parallaxes, radial velocities, and proper motions, suggesting their association.

\begin{figure}
    \centering
    \includegraphics{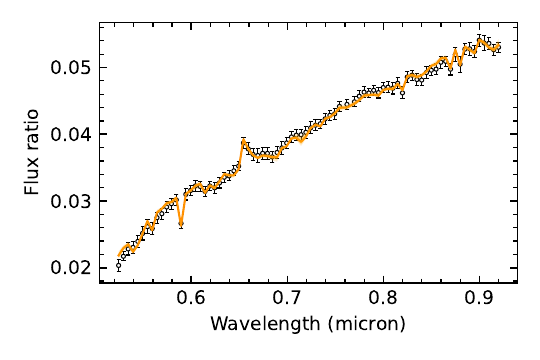}
    \caption{Narrowband flux ratios between the companion and HAT-P-41. The errorbars are the observed flux ratios, while the orange line is the best-fit curve using the PHOENIX model spectra.}
    \label{fig:flux_ratio}
\end{figure}

To correct the flux contamination from the companion star, we performed a supplementary observation with 10 exposures during OB-3, where the slit was aligned to the direction from HAT-P-41 to the companion star, so that the spectra of the two stars could be spatially resolved. The distance between the PSF peaks of the two stars is $\sim$3.5\arcsec (14 pixels). To separate the two stellar spectra, we first subtracted the mirrored non-contaminated side of the target's PSF from the other side to obtain the companion's spectrum. Since strong residuals existed at the peak location of the target, we fitted a Gaussian function to target's original PSF and made a mask for regions within 4$\sigma$ from the peak location. We then repaired the masked region by mirroring the non-contaminated part of the companion's PSF on the other side. The target's spectrum was then obtained by subtracting the extracted companion's spectrum from the total spectrum. The companion-to-host flux ratio in each passband was then calculated based on the extracted stellar spectra. The broadband flux ratio (521.8 -- 922.4 nm excluding the oxygen A band) was measured to be $0.03885 \pm 0.00083$. The narrowband flux ratios are shown in Fig. \ref{fig:flux_ratio}.

When fitting the light curve in each passband, the flux dilution effect was corrected by
\begin{equation}
    f(t) = \frac{f^*(t) +\mathcal{F}}{1+\mathcal{F}},
\end{equation}
where $\mathcal{F}$ is the companion-to-host flux ratio and $f^*(t)$ is the light curve model without flux dilution.  

We further refined the constraints on the physical parameters of the companion star by fitting the observed flux ratios with PHOENIX model spectra \citep{2013A&A...553A...6H} based on the stellar parameters from \cite{2012AJ....144..139H} and parallaxes measurements from the Gaia data \citep{2016A&A...595A...1G, 2023A&A...674A...1G}. This method was detailed in \cite{2016A&A...594A..65N} and \cite{2023A&A...675A..62J}. Figure \ref{fig:flux_ratio} shows the fitting results of the flux ratios. The derived effective temperature of $4498^{+28}_{-33}$~K and radius of $0.708 \pm 0.022$~$R_\odot$ suggest a K-type companion star. However, to accurately determine its gravity, mass, and spectral type, further high-resolution spectroscopic measurements are necessary.

\subsection{Gaussian process modeling}

The light curve systematic noise was fitted with the multi-input Gaussian process (GP) regression implemented by the Python package \texttt{George} \citep{2015ITPAM..38..252A}. The GP models were applied to all the broadband and narrowband light curves individually. We adopted the 3/2-order Mat\'ern kernel function $k(r)$ with multi-input variables (time $t$, seeing\footnote{Measured with the full width at half maximum of the stellar PSF fitted by a Gaussian function along the spatial direction at the central wavelength of $\sim$750~nm.} $w$, and pixel drift\footnote{ Measured with the pixel drift of the stellar PSF fitted by a Gaussian function along the spatial direction at the central wavelength of $\sim$750~nm.} $p$) for GP modeling:
\begin{equation}
\begin{split}
    k(r) &= \left(1+\sqrt{3r^2}\right) \exp{\left(-\sqrt{3r^2}\right)}, \\
    k_{\rm total} (\bm{r}) &= \sigma_k^2 \cdot k\left(\frac{\Delta t}{\ell_t}\right)
        \cdot k\left(\frac{\Delta w}{\ell_w}\right)
        \cdot k\left(\frac{\Delta p}{\ell_p}\right),         
\end{split}
\end{equation}
where $\sigma_k^2$ is the variance of the kernel, $\bm{r}$ is the distance between two input vectors $(t, w, p)$ and $(t+\Delta t, w+\Delta w, p+\Delta p)$ normalized by the corresponding length scales $\ell_t$, $\ell_w$, and $\ell_p$. In addition, the uncertainties of the normalized light curves were estimated as photon-dominated noise and usually underestimated the total white noise. To compensate for this, a jitter variance of $\sigma_n^2$ was added to the diagonal of the covariance matrix ($\bm{r}=\bm{0}$). Therefore, the free GP parameters in GP regression are $\sigma_n$, $\sigma_k$, $\ell_t$, $\ell_w$, and $\ell_p$ with log-uniform priors. 

\subsection{Fitting results} \label{sect:bayesian}

\begin{figure*} 
    \centering
    \includegraphics{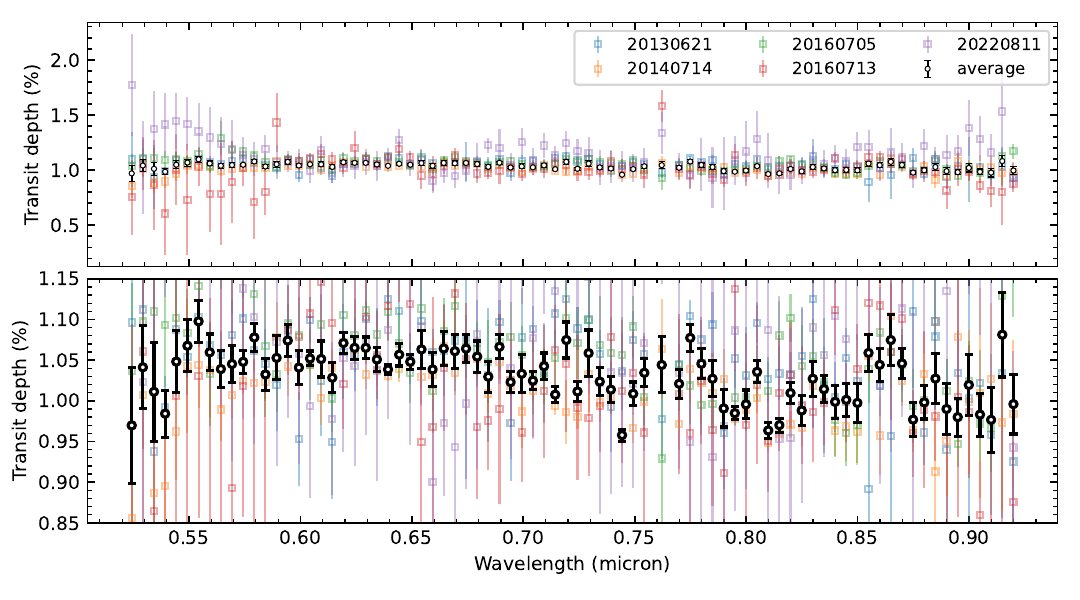}
    \caption{Transmission spectra of HAT-P-41b derived from the five transit observations with OSIRIS R1000R. {\it Upper panel:} Five individual transmission spectra (colored error bars) and their weighted means (black error bars). {\it Lower panel:} Same as the upper panel, but zoomed in for clarity.}
    \label{fig:averaged_spectrum}
\end{figure*}

We used the nested sampling algorithm implemented by \texttt{PyMultiNest} \citep{2014A&A...564A.125B} to obtain posterior estimates and model evidence (see also \citealt{2008MNRAS.384..449F} and \citealt{2009MNRAS.398.1601F} for more information on the \texttt{MULTINEST} algorithm). We used 1000 live points and a sampling efficiency of 0.3 in each run to reach a log-evidence precision of $\sim$0.1 when fitting transit light curves. The fitting results of the broadband light curves are shown in Fig. \ref{fig:white_light}, while those of the spectroscopic light curve can be found at \texttt{ScienceDB}\footnote{\url{https://doi.org/10.57760/sciencedb.11105}}. The posterior estimates of the central transit epochs are listed in Table \ref{table:transit_params}. As shown in Fig. \ref{fig:averaged_spectrum}, the five transmission spectra observed with GTC OSIRIS showed good consistency in most wavebands, while discrepancies in the bluer and the redder ends are due to very strong systematics in the spectroscopic light curves, which were compensated by the very large uncertainties. While the removal of the common-mode noise helps to reduce the amplitudes of light curve systematics and increase the precision of spectroscopic transit depths across most wavebands, the residual wavelength-dependent systematics could still result in the wavelength dependency of transit depth uncertainties. Given the low activity of the host star \citep{2012AJ....144..139H, 2021AJ....161...51S}, the subsequent retrieval analysis was performed using the weighted average of the five transmission spectra.

\begin{table}
\renewcommand\arraystretch{1.2}
\caption{Posterior estimates of the central transit time of HAT-P-41b.} 
\label{table:transit_params} 
\centering 
\begin{tabular}{c c c } 
\hline\hline 
\# & Date & $T_{\rm c}~({\rm BJD_{TDB}})$  \\ 
\hline 
OB-1 & 2013-06-21 & 
    $2456465.59147^{+0.00031}_{-0.00034}$  \\
    
OB-2 & 2014-07-14 &
    $2456853.53273^{+0.00119}_{-0.00143}$ \\

OB-3 & 2016-07-05 &
    $2457575.53814^{+0.00027}_{-0.00030}$ \\

OB-4 & 2016-07-13 &
    $2457583.61970^{+0.00041}_{-0.00038}$ \\

OB-5 & 2022-08-11 &
    $2459803.51788^{+0.00061}_{-0.00061}$ \\

\hline 
\end{tabular}
\end{table}

\section{Atmospheric retrievals}
\label{sect:retrievals}

\subsection{Model setup}

We used the Python package \texttt{petitRADTRANS} \citep{2019A&A...627A..67M} to model the transmission spectrum of HAT-P-41b. A one-dimensional isothermal atmosphere was assumed with a pressure range from $10^{-6}$ to $10^2$~bar and 100 uniform layers in logarithmic space. We included 12 chemical species (Na, K, MgO, AlO, TiO, VO, MgH, AlH, CaH, CrH, FeH, and $\rm H_2O$) that could show significant spectral features in the OSIRIS wavelength range, and four additional species ($\rm NH_3$, $\rm CH_4$, CO, and $\rm CO_2$) when retrieving the optical-to-near-infrared (NIR) joint transmission spectrum. We assumed free mass fractions $m_i$ for each chemical species $i$ and used $\rm H_2$ and $\rm He$ as filler gases to keep the total mass fraction unity with a fixed $\rm He/H_2$ mass ratio of 0.3. Rayleigh-like scattering and collision-induced absorption of $\rm H_2$ and He were also included in the model, with the amplitude of the scattering features multiplied by a factor of $A_{\rm s}$. Additionally, we assumed a uniform opaque cloud cover with a cloud top pressure of $P_{\rm c}$, below which all transmitted light would be blocked. The forward model was originally computed with a spectral resolution ($\lambda/\Delta\lambda$) of 1000 then wavelength binned to the data passbands.

To infer the possible species from the spectral features, we removed one chemical species at a time from the free chemistry model and computed the difference of the log-evidence with and without that species, which is equivalent to the logarithmic Bayes factor. We interpret the Bayes factors using the criteria proposed by \cite{Kass1995}, where the evidence is strong if $|\Delta \ln \mathcal{Z}|\ge5$, moderate if $3\le|\Delta \ln \mathcal{Z}|<5$, weak if $1\le|\Delta \ln \mathcal{Z}|<3$, and inconclusive if $|\Delta \ln \mathcal{Z}|<1$. The Bayesian evidence and posterior estimates were obtained using the nested sampling algorithm in the same way as introduced in Sect. \ref{sect:bayesian}, but sampled with fewer live points ($N=500$) due to the very high computational cost, which still guaranteed $\sim$150\,000 likelihood evaluations and $\sim$3500 posterior samples.

\subsection{OSIRIS transmission spectrum}

\begin{table}
\caption{Bayesian model comparison of the chemical species retrieved with the OSIRIS transmission spectrum.} 
\label{table:evidence} 
\centering 
\begin{tabular}{l c c} 
\hline\hline 
Species & $\Delta \ln \mathcal{Z}$ \tablefootmark{a} & Inferences \\ 
\hline 
Na & 0.11 & inconclusive \\
K & -0.46 & inconclusive  \\

MgO & 0.22 & inconclusive  \\
AlO & 0.37 & inconclusive  \\
TiO & \textbf{21.02} & strong \\
VO & -0.58 & inconclusive \\

MgH & \textbf{2.32} & weak \\
AlH & -0.34 & inconclusive \\
CaH & 0.09 & inconclusive \\
CrH & \textbf{3.73} & moderate \\
FeH & -0.4 & inconclusive \\

H$_2$O & 0.15 & inconclusive  \\

\hline 
\end{tabular}
\tablefoot{
    \tablefoottext{a}{Logarithmic Bayesian evidence ($\ln \mathcal{Z}$) of the full retrieval subtracting that of the hypothesis excluding one chemical species, where the uncertainty of $\ln \mathcal{Z}$ is $\sim$0.15. The values with boldface indicate |$\Delta \ln \mathcal{Z}|\ge1$.}
} 

\end{table}

\begin{figure*}
    \centering
    \includegraphics{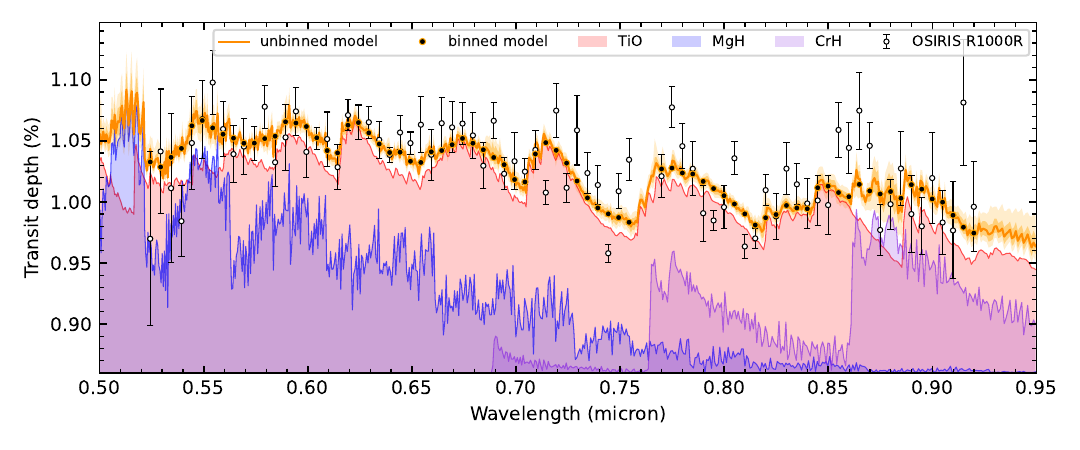}
    \caption{Retrieved transmission spectrum of HAT-P-41b observed with GTC OSIRIS. The error bars are the averaged OSIRIS transmission spectrum. The orange lines are the median and $2\sigma$ credible intervals of the posterior models. The black dots are the posterior model after wavelength binning to the observational passbands. The shaded regions below the posterior model indicate the reference models with only TiO (red), MgH (blue), or CrH (purple), respectively.}
    \label{fig:retrieval_vis}
\end{figure*}

We first computed the full atmospheric retrieval including all 12 chemical species (Fig. \ref{fig:retrieval_vis}). The posterior joint distributions of the atmospheric parameters can be found at \texttt{ScienceDB}\footnote{\url{https://doi.org/10.57760/sciencedb.11105}}. The retrieved model suggested a clear atmosphere ($\log_{10} P_{\rm c}=-0.74^{+1.64}_{-1.53}$~$\log_{10} {\rm bar}$) with a temperature of $2447^{+299}_{-310}$~K, higher than the planetary equilibrium temperature \citep[$1941\pm38~{\rm K}$,][]{2012AJ....144..139H}. The Bayesian evidence of the full retrieval ($\ln \mathcal{Z}=501.29 \pm 0.16$) was much stronger than that of a flat model ($\ln \mathcal{Z}=402.18 \pm 0.12$) or a pure scattering model ($\ln \mathcal{Z}=467.70 \pm 0.14$), suggesting significant spectral features in the OSIRIS transmission spectrum.  These spectral features can mostly be characterized by the absorption signatures of TiO, while near the blue and red ends some excess absorption could be characterized by the absorption of MgH and CrH, respectively. 

We then performed model comparison based on Bayesian evidence to validate the detection of each chemical species. We removed each chemical species from the full retrieval and estimated the Bayesian evidence for a model hypothesis excluding that species. If the new Bayesian evidence is substantially less than the Bayesian evidence of the full retrieval, it indicates that the absorption features of the missing species cannot be counteracted by other species, implying the need for that species to characterize the transmission spectrum. According to Table \ref{table:evidence}, there is very strong evidence for the presence of spectral features of TiO ($\Delta\ln \mathcal{Z}=21.02$), moderate evidence for that of CrH ($\Delta\ln \mathcal{Z}=3.73$), and weak evidence for that of MgH ($\Delta\ln \mathcal{Z}=2.32$) in the OSIRIS transmission spectrum, which can be approximated as a detection significance of 6.8$\sigma$ for TiO, 3.2$\sigma$ for CrH, and 2.7$\sigma$ for MgH \citep{Sellke2001, 2013ApJ...778..153B}. We found no evidence for the absorption signatures of the alkali metals or other metal oxides and hydrides. 

The absorption features of TiO in the atmosphere of HAT-P-41b have been previously detected by HST STIS measurements \citep{2021AJ....161...51S}, while those of CrH and MgH were discovered for the first time in this work. \cite{2021AJ....161...51S} compared the results of two transmission spectrum models: the equilibrium chemistry models of \texttt{PLATON} and the free chemistry models of \texttt{AURA}. They found suggestive evidence for the optical absorbers and decisive evidence for water vapor in the 0.3 -- 5.0 $\mu$m transmission spectrum of HAT-P-41b. While the free chemistry retrieval from \texttt{AURA} supports the presence of absorption features of Na I and AlO in the optical band, the results from \texttt{PLATON} favor the absorption of TiO and VO. However, we did not detect the absorption features of Na I and AlO in the OSIRIS transmission spectrum using the \texttt{petitRADTRANS} free chemistry model. Instead, we observed significantly higher chemical abundances of TiO compared to VO, which is different from the predictions of the equilibrium chemical model. Additionally, we found preliminary evidence for MgH and CrH, which are not included in the equilibrium chemistry model. 

\section{Discussion and conclusion} \label{sect:dis_and_con}

\begin{figure*}
    \centering
    \includegraphics{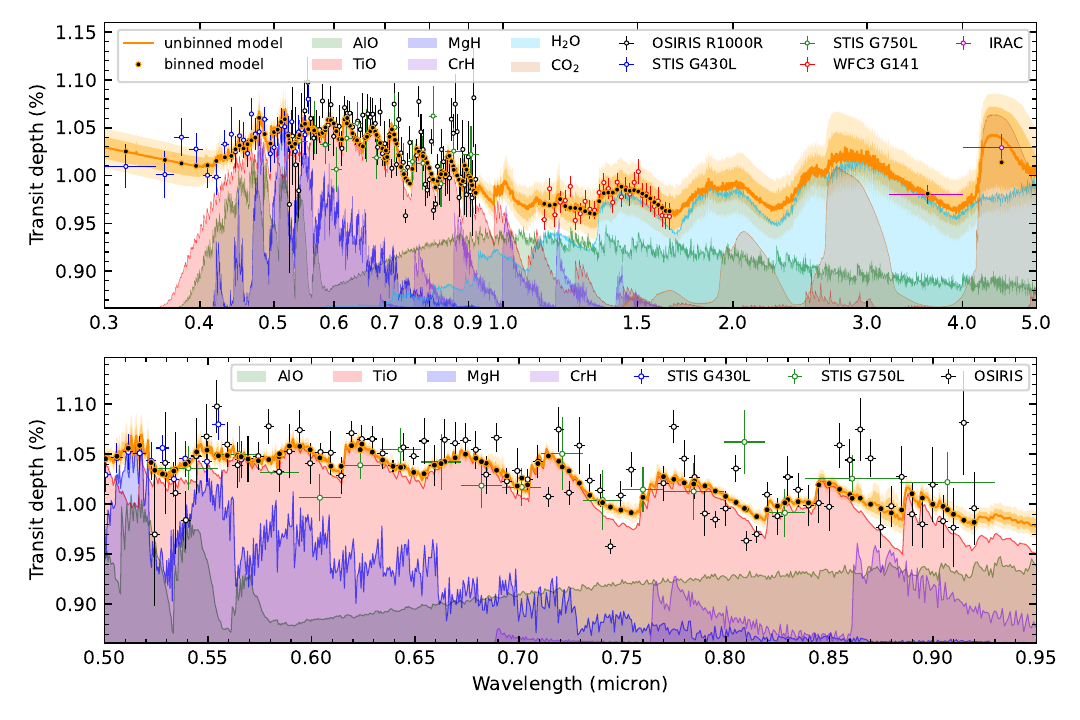}
    \caption{Retrieval results of HAT-P-41b using the optical-to-NIR joint transmission spectra. {\it Upper panel}: The error bars are the transmission spectra measured with different instruments, where the median instrumental offsets have been applied ($-4$~ppm for STIS, 758~ppm for G141, and 388~ppm for IRAC). The median and $2\sigma$ credible interval of the posterior models are shown in orange lines. The black dots are the median model after wavelength binning. The shaded regions below the posterior model indicate the reference models with only AlO (green), TiO (red), MgH (deep blue), CrH (purple), $\rm H_2O$ (sky blue), and $\rm CO_2$ (brown) respectively. {\it Lower panel}: Same as the upper panel, but zoomed into the wavelength range of the OSIRIS transmission spectrum.}
    \label{fig:retrieval_nir}
\end{figure*}

\begin{figure*}
    \centering
    \includegraphics{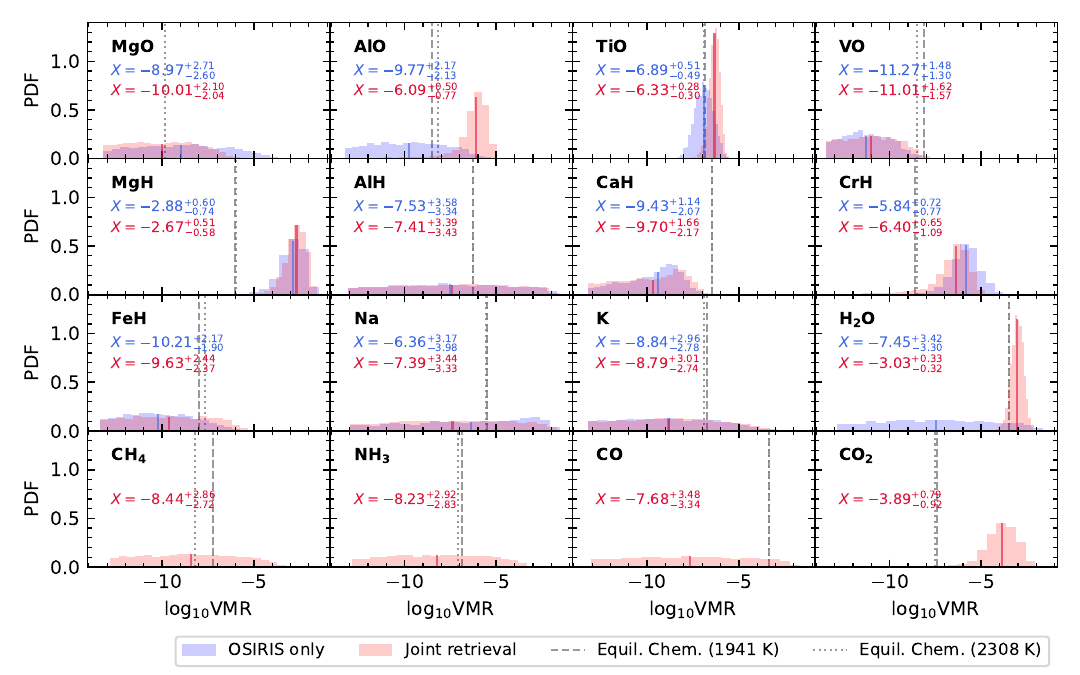}
    \caption{Posterior distributions of the VMRs of chemical species constrained by the OSIRIS transmission spectrum (blue) and the optical-to-NIR joint spectrum (red). The symbol $X$ denotes the logarithmic volume mixing ratio $\rm \log_{10}VMR$ with the corresponding colors. The gray lines indicate the VMRs calculated using \texttt{GGChem} equilibrium chemistry \citep{2018A&A...614A...1W} assuming solar abundances at a pressure of 1~bar, where two planetary equilibrium temperatures were considered: 1941~K (dashed, assuming uniform heat redistribution) and 2308~K (dotted, assuming dayside heat redistribution only). }
    \label{fig:dist_abundances}
\end{figure*}

To precisely constrain the atmospheric chemical abundances, we performed a joint retrieval by combining the OSIRIS transmission spectrum with the 0.3 -- 5.0 $\mu$m transmission spectrum presented in \cite{2021AJ....161...51S}. To account for possible transit depth offsets due to different instruments and/or light curve analysis methods, three arbitrary offsets were added as free parameters with a uniform prior of $\mathcal{U}(-1000, 1000)$~ppm to the HST STIS, HST WFC3 and Spitzer IRAC transmission spectra, separately. As shown in Fig. \ref{fig:retrieval_nir}, the retrieved model from the joint transmission spectra exhibits strong absorption signatures of TiO, $\rm H_2O$, and $\rm CO_2$, and tentative signatures of AlO, MgH, and CrH, which are in good agreement with the results from the OSIRIS transmission spectrum in the optical wavelength range. 

Figure \ref{fig:dist_abundances} presents the posterior volume mixing ratios (VMRs\footnote{Volume mixing ratio ${\rm VMR} = m_i \cdot  \mu / \mu_i$, where $m_i$ and $\mu_i$ are the mass fraction and the particle mass of the species $i$, and $\mu$ is the atmospheric mean molecular weight.}) of each chemical species obtained from the atmospheric retrievals, along with the VMRs predicted by the equilibrium chemistry model computed with \texttt{GGChem} \citep{2018A&A...614A...1W} assuming solar abundances. The chemical abundances constrained by the OSIRIS transmission spectrum alone are consistent with those constrained by the joint spectrum, except for AlO and $\rm H_2O$. Based on the joint retrieval, we obtained precise measurements of the abundances of AlO ($-6.09^{+0.50}_{-0.77}$~dex), TiO ($-6.33^{+0.28}_{-0.30}$~dex), MgH ($-2.67^{+0.51}_{-0.58}$~dex), CrH ($-6.40^{+0.65}_{-1.09}$~dex), $\rm H_2O$ ($-3.03^{+0.33}_{-0.32}$~dex), and CO ($-3.89^{+0.79}_{-0.92}$~dex). \cite{2021AJ....161...51S} also reported the detection of AlO in the optical wavelength range. We note that the main spectral features of AlO are from 0.40 to 0.55 microns and thus the measurement on AlO was mainly contributed by the STIS G430L spectrum rather than the OSIRIS R1000R spectrum. In addition, the low abundances of MgO, VO, CaH, and FeH disfavored the presence of these species in the terminator atmosphere of HAT-P-41b. Comparing the retrieved VMRs with the equilibrium chemistry model assuming solar abundances, we found super-solar abundances for all well-constrained chemical species in the joint retrieval (AlO, TiO, MgH, CrH, $\rm H_2O$, and $\rm CO_2$). Although the detections of MgH and CrH were tentative, there is strong evidence for a super-solar TiO abundance (and a super-solar $\rm H_2O$ abundance when considering the NIR data), supporting a metal-rich atmosphere for HAT-P-41b.

The high metallicity of the host star ($\rm [Fe/H]=0.21 \pm 0.10$; \citealt{2012AJ....144..139H}) provides the conditions for the formation of a metal-rich atmosphere for HAT-P-41b. Considering that the equilibrium temperature of HAT-P-41b ($T_{\rm eq}=1941 \pm 38$~K) is in the empirical transition zone between hot and ultra-hot planets around 2150 K \citep{2022A&A...662A.101S}, its atmospheric composition may contain both characteristics in the low-temperature and high-temperature regimes from a statistical point of view. Therefore, aside from the detected $\rm H_2O$, $\rm CO_2$, and refractory molecules, atomic species such as H, Fe, Mg, and alkali metals may also be expected in the atmosphere of HAT-P-41b in future high-resolution observations. 

The first tentative evidence for CrH was found in the atmosphere of WASP-31b, a hot Jupiter with an equilibrium temperature $\sim$370~K cooler than HAT-P-41b, based on the observations of HST STIS G750L \citep{2021A&A...646A..17B}. The presence of CrH in WASP-31b was recently confirmed by high-resolution transmission spectroscopy \citep{2023ApJ...953L..19F}. With the addition of HAT-P-41b that potentially bares CrH in the atmosphere, future studies could target at the formation of Cr-bearing clouds in the planetary atmosphere \citep{2006asup.book....1L, 2012ApJ...756..172M}.

There are other hot Jupiters with temperatures, masses, radii, and host star parameters similar to HAT-P-41b, for example, CoRoT-1b \citep{2008A&A...482L..17B}, HAT-P-7b \citep{2008ApJ...680.1450P}, HAT-P-32b \citep{2011ApJ...742...59H}, HAT-P-65b \citep{2016AJ....152..182H}, and WASP-19b \citep{2010ApJ...708..224H}. Observations of these planets show that hot Jupiters with transition-zone equilibrium temperatures orbiting F-G type stars can exhibit rich spectral features of both volatile species (e.g., $\rm H_2O$) and refractory species (e.g., TiO) if not obscured by high-altitude clouds or hazes \citep{2013MNRAS.434.3252H, 2017AJ....154...39D, 2017Natur.549..238S,  2018AJ....155..156T, 2021ApJ...913L..16C, 2021MNRAS.505..435S, 2022A&A...657A...6C, 2022AJ....164...19G, 2022ApJS..260....3C, 2023RAA....23b5018L, 2023A&A...675A..62J}. The day-night and morning-evening differences could allow $\rm H_2$ and $\rm H_2O$ to be dissociated into $\rm H^-$ in the hot day-side and evening-side atmospheres, while recombining into water molecules or other metal hydrides in the cooler night-side and morning-side atmospheres \citep{2018A&A...617A.110P}. 
Our preliminary detection of MgH and CrH in the terminator atmosphere sheds lights on the presence of $\rm H^-$ in the hot day-side atmosphere. We speculate that the emission features of the day-side atmosphere might be obscured by the continuum absorption of $\rm H^-$, resulting in an emission spectrum with weak features and thus tending toward an isothermal profile, which would explain why the thermal inversion based on the secondary eclipse measurements from HST and Spitzer was inconclusive \citep{2022AJ....163..190F}.

In this study, we analyzed the optical transmission spectrum of HAT-P-41b using data from five transits observed by GTC OSIRIS. Our free-chemistry atmospheric retrievals and model comparisons provided strong evidence for the presence of TiO in the atmosphere of HAT-P-41b, and we also tentatively detected the signatures of MgH and CrH. Our joint retrieval analysis combining GTC, HST, and Spitzer transmission spectra yielded results consistent with the analysis of the OSIRIS data and provided precise constraints on the abundances of TiO, MgH, CrH, $\rm H_2O$, and $\rm CO_2$. The detection of abundant optical absorbers in HAT-P-41b’s high-temperature atmosphere suggests the possibility of a thermal inversion in its upper atmosphere. However, current thermal emission spectra have not provided decisive evidence for this inversion \citep{2021NatAs...5.1224M, 2022AJ....163..190F, 2022ApJS..260....3C}, which may be due to the presence of H$^-$ obscuring other emission features. Future observations with broader wavelength coverage or higher spectral resolution, such as those to be obtained by the James Webb Space Telescope (JWST) or other large ground-based telescopes, are needed to further characterize the day-side atmosphere of HAT-P-41b and confirm the signatures of metal hydrides.

\begin{acknowledgements}
We thank the anonymous referee for their valuable comments and suggestions.
G.C. acknowledges the support by the National Natural Science Foundation of China (grant Nos. 12122308, 42075122), the B-type Strategic Priority Program of the Chinese Academy of Sciences (grant No. XDB41000000), Youth Innovation Promotion Association CAS (2021315), and the Minor Planet Foundation of the Purple Mountain Observatory. This work is based on observations made with the Gran Telescopio Canarias installed at the Spanish Observatorio del Roque de los Muchachos of the Instituto de Astrofísica de Canarias on the island of La Palma, and on data from the GTC Archive at CAB (CSIC-INTA). The GTC Archive is part of the Spanish Virtual Observatory project funded by MCIN/AEI/10.13039/501100011033 through grant PID2020-112949GB-I00. 
This work has made use of data from the European Space Agency (ESA) mission {\it Gaia} (\url{https://www.cosmos.esa.int/gaia}), processed by the {\it Gaia} Data Processing and Analysis Consortium (DPAC, \url{https://www.cosmos.esa.int/web/gaia/dpac/consortium}). Funding for the DPAC has been provided by national institutions, in particular the institutions participating in the {\it Gaia} Multilateral Agreement.
\end{acknowledgements}

\bibliographystyle{aa}
\bibliography{references}

\onecolumn

\begin{appendix}

\section{Supplementary figures and tables}

\begin{table*}[!h]
\caption{Observation summary for HAT-P-41b.} 
\label{table:observation} 
\centering 
\begin{tabular}{c c c c c c c c c } 
\hline\hline 
\# & Proposal ID & Date (UT) & 
\makecell{Slit \\ ($''$)}& \makecell{Readout \\ (kHz)} & 
\makecell{Exp. time \\ (s)} & Useful exposures & 
\makecell{Airmass \\ (start -- min -- end)} & 
\makecell{Seeing\tablefootmark{a} \\ (95\% interval)}\\ 
\hline 

OB-1 & GTC52-13A\tablefootmark{b} & 2013-06-21 & 12 & 200 & 8.0 & 426 & 
1.71 -- 1.09 -- 1.23 & $1.07''$ -- $2.34''$ \\ 

OB-2 & GTC9-14A\tablefootmark{b} & 2014-07-14 & 40 & 200 & 7.0 & 668 & 
2.27 -- 1.10 -- 1.18 & $0.69''$ -- $1.60''$\\

OB-3 & GTC65-16A\tablefootmark{c} & 2016-07-05 & 12 & 500 & 7.5 & 987 & 
1.74 -- 1.09 -- 1.33 & $0.84''$ -- $1.32''$ \\

OB-4 & GTC65-16A\tablefootmark{c} & 2016-07-13 & 12 & 500 & 20.0 & 707 & 
1.39 -- 1.09 -- 2.18 & $1.37''$ -- $2.24''$\\

OB-5 & GTC90-22A\tablefootmark{d} & 2022-08-11 & 40 & 200 & 10.0 & 602 & 
1.35 -- 1.10 -- 2.43 & $0.94''$ -- $1.73''$ \\

\hline 
\end{tabular}
\tablefoot{
    The OSIRIS instrument was positioned at the Nasmyth-B focus of GTC for the first four observations, and at the Cassegrain focus for OB-5.
    \tablefoottext{a}{Seeing is measured with the full width at half maximum of the stellar PSF along the spatial direction at the central wavelength of $\sim$750~nm.}
    \tablefoottext{b}{PI: E. Pall\'e.}
    \tablefoottext{c}{PI: R. Alonso.}
    \tablefoottext{d}{PI: F. Murgas.}
    
} 
\end{table*}

\begin{table*}[h]
\caption{Measurements of the HAT-P-41 system and updated constraints on its companion.} 
\label{table:companion} 
\centering 
\renewcommand\arraystretch{1.2}
\begin{tabular}{l c c c} 
\hline\hline 
Parameters & HAT-P-41 & Companion & References\\ 
\hline 
Right ascension (deg) & 297.322651868 & 297.322580691 & [1] \\
Declination (deg) & 4.672411376 & 4.671410231 & [1] \\
Parallax (mas) & $2.85 \pm 0.02$ & $2.91 \pm 0.03$ & [1] \\
Radial velocity (km/s) & $32.54 \pm 0.47$ & $32.63 \pm 3.81$ & [1] \\
Proper motion (mas/yr) & 7.30 & 7.65 & [1] \\
\\
Effective temperature (K) & $6390 \pm 100$ & $4498^{+28}_{-33}$  & [2], this work \\
Metallicity (dex) & $0.21 \pm 0.10$ & $0.41 ^{+0.13}_{-0.17}$ & [2], this work \\
Gravity (cgs) & $4.14 \pm 0.02$ & $4.30 ^{+0.19}_{-0.20}$ & [2], this work \\
Mass ($M_\sun$) & $1.418 \pm 0.047$ & $0.36 ^{+0.19}_{-0.19}$ & [2], this work \\
Radius ($R_\sun$) & $1.683 ^{+0.058}_{-0.036} $ & $0.708 ^{+0.022}_{-0.022}$ & [2], this work \\
\hline 
\end{tabular}

\tablebib{
    [1] \cite{2016A&A...595A...1G, 2023A&A...674A...1G}
    [2] \cite{2012AJ....144..139H}
}
    
\end{table*}

\begin{figure*}[h]
    \centering
    \includegraphics{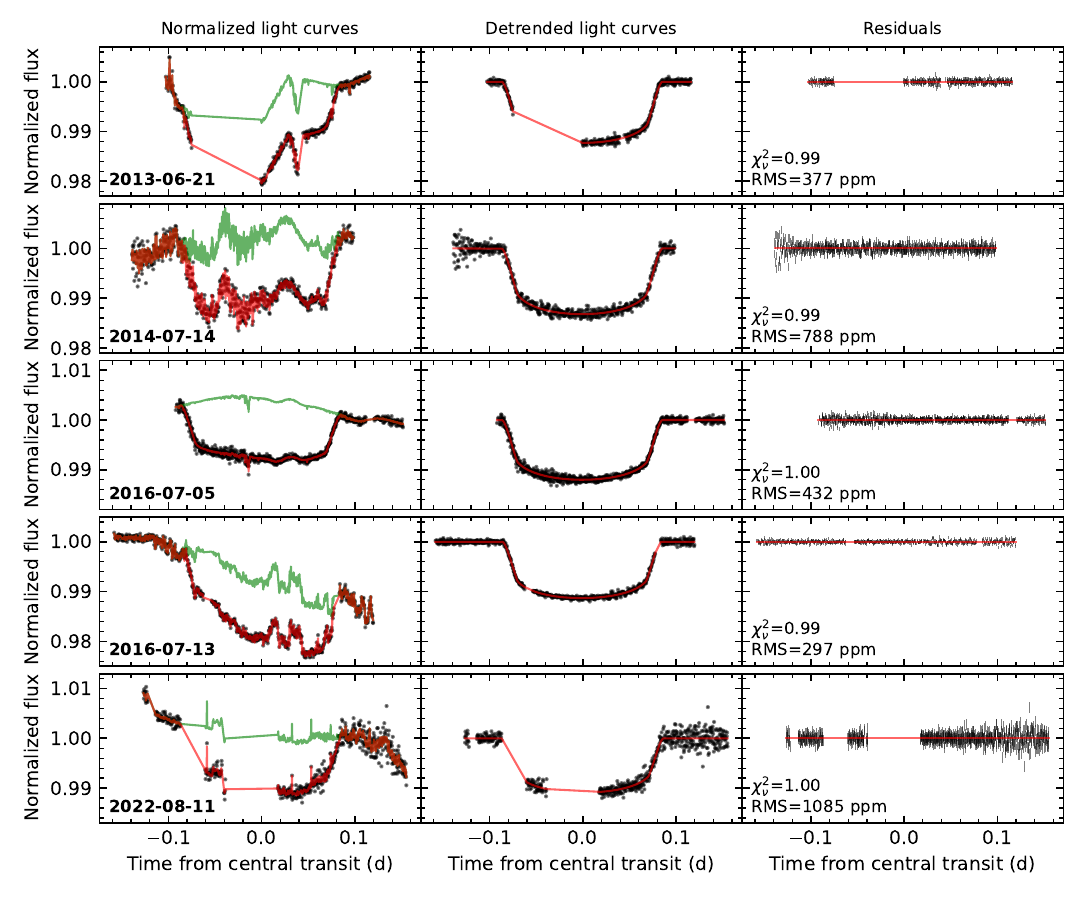}
    \caption{Fitting results of the broadband transit light curves of HAT-P-41b. Each row corresponds to one transit observation. \textit{Left column}: the normalized fluxes (black error bars), the best-fit models (red), and the systematic noise (green, also the common-mode systematics). \textit{Middle column}: the detrended light curves and the best-fit transit models. \textit{Right column}: the residuals.}
    \label{fig:white_light}
\end{figure*}

\end{appendix}

\end{document}